# Imagining the Musical Metaverse: a study on the needs and creative practices of composers and musicians

**Authors:** Alberto Boem, Matteo Tomasetti, Alessio Gabriele, Agostino di Scipio, Alessandra Micalizzi, Luca Turchet

**Abstract**

This study explores user needs for the Musical Metaverse (MM) through a series of workshops with electroacoustic composers, classical musicians, and music producers. Using a design approach based on the prompt 'as if by magic,' participants were invited to reflect on how the MM could impact their practice in composition, performance, and education domains. While groups maintain distinct priorities based on their roles, they share interests in educational applications and creative learning environments. Key themes include preference for mixed reality over purely virtual environments, virtual space as a creative paradigm, and tensions between democratizing tools and maintaining authentic musical experiences. Education emerged as the most promising initial use case, particularly for understanding complex musical concepts. However, participants expressed skepticism about fully virtual performances, emphasizing physical connections to instruments and spaces. Results suggest MM development should enhance rather than replace traditional practices while supporting embodied creativity.

## 1. Introduction

In recent years, interest in the concept of the Metaverse has surged once again, especially in fields such as engineering, computer science, and telecommunications (El Saddik et al. 2024; Ritterbusch and Teichmann 2023; Wang, Ning et al. 2023). Broadly, the Metaverse can be described as a vision of technology-mediated collaborative environments that bridge the gap between the physical and virtual worlds (Ball 2024). The term 'Metaverse' first appeared in Neal Stephenson's 1992 novel 'Snow Crash' (Stephenson 1994), initially referring to a 3D virtual world where people exist as avatars and live alternative lives. Since then, this fictional concept has gained the attention of the emerging industry around the World Wide Web and immersive technologies, leading to the emergence of platforms like 'Active Worlds' in the 1990s and 'Second Life' around 2003, thanks also to the attention on shared virtual environments (Churchill and Snowdon 1998; Waters and Barrus 1997).

Since then, several platforms have emerged (i.e. Decentraland, Somnium Space, Horizon Worlds), and research has started to explore both the technical aspects (Cheng, Wu, Chen et al. 2022; Cheng, Wu, Varvello et al. 2022; Sai, Garg, and Chamola 2024; Tang et al. 2023) and ethical challenges of the Metaverse (Benjamins, Viñuela, and Alonso 2023; Munn and Weijers 2023; Spence 2008). However, a paper from 2022 identified 260 scientific publications, each offering its own definition of the Metaverse (Park and Kim 2022). When combined with marketing and media attention, especially after the rebranding of Facebook into Meta, this definitional diversity has spurred controversy among both the general public and expert audiences, and leaving the term increasingly ambiguous.

Even with this lack of consensus, some characteristics of the Metaverse have been identified. The Metaverse should allow geographically dispersed users to interact in shared, interoperable, and persistent immersive environments through the use of digital media. It is characterized as a virtual space connected to or reflecting aspects of the physical world, supporting and facilitating social interactions through multisensory and immersive experiences. Notably, the Metaverse is not merely a new term for Augmented, Mixed, and Virtual Reality (we will refer to these from now on as AR, MR, and VR) or the more comprehensive eXtended Reality (XR) or 'Spatial Computing' (Shekhar, Feiner, and Aref 2015), but rather a convergence of various technologies including the Internet of Things, edge computing, high-speed internet communication, and digital twins. While the Metaverse concept has been explored in various domains such as tourism (Gursoy, Malodia, and Dhir. 2022), education (Lin et al. 2022), security (Wang, Su et al. 2023), and healthcare (Chengoden et al. 2023), music remains an underexplored area of application.

In response, the vision of the Musical Metaverse (MM) has been proposed to expand the potential of this concept into the realm of music (Turchet 2023). The MM is defined as the component of the Metaverse focusing specifically on musical activities. This emerging space combines elements of augmented realities and virtual worlds with the goal of changing how we create, share, and enjoy music. The MM concept is built on the convergence of several technologies, including Musical XR (Turchet, Hamilton, and Çamci 2021), Networked Music Performance (NMP) (Renaud, Carôt, and Rebelo 2007), and the Internet of Musical Things (IoMusT) (Turchet et al. 2018). Musical XR sits at the intersection of interactive music and immersive technologies such as AR, MR, and VR. IoMusT aims to expand the Internet of Things (IoT) paradigm to include the creation and use of networked musical devices. This would foster innovative interactions among musicians, audiences, and other musical stakeholders. NMP systems form a central part of IoMusT, and they allow musicians in different locations to play together in real-time over a network.

The MM is envisioned to support a wide range of musical activities beyond just concert experiences but is still in its early stages of development. Despite growing interest from the research community, a proper and functional MM does not yet exist. At present, users can only experience independent immersive social environments or prototypes dedicated to selected musical activities and practices. On the other hand, given its potential social and economic impact, technology companies and industries are actively pursuing the development of a fully operational Metaverse (Dwivedi et al. 2022).

However, existing musical platforms lack the precision and reliability needed by professional musicians, particularly for collaborative performances and concerts. In this context, audience participation in immersive environments has received significant attention (Onderdijk et al. 2023; Park, Choi, and Lee 2024; Ppali et al. 2024).

Some of the other main challenges currently addressed by the research community are mostly technical (Boem, Tomasetti, and Turchet 2025). They include: (a) Latency, jitter, and bandwidth limitations in creating realistic experiences for multiple musicians connected online; (b) Limited use of audio in the Metaverse, as it is primarily used for enhancing immersion rather than capturing the nuances of music-making; (c) The need for domain-specific tools to achieve high levels of audio quality and interaction.

Despite these challenges, the MM aims to be more than just a collection of technologies. The MM should support social presence and interactions (Oh, Bailenson, and Welch 2018), enabling users to inhabit and share remote environments that facilitate collaboration and communication (Schroeder et al. 2001). Due to its social nature and demanding requirements for coordination, timing, and expression, music applications can play a crucial role in addressing the challenges of developing the Metaverse as a whole.

Although interest in the MM has grown in recent years leading to the establishment of international scientific gatherings explicitly dedicated to the topic, both the XR and music research communities have yet to explore it extensively. Only a small number of publications have specifically addressed the MM from the points of view of experts and stakeholders (Boem, Tomasetti, and Turchet 2024; Bruns et al. 2024; Loveridge 2024).

To promote and advance the development of the MM, it is crucial to define elements such as user needs and expectations, which helps in to shaping design requirements and the user experience. Our study aims to explore future applications and requirements helpful in creating an MM centred on the needs of musicians and sound creators, particularly electroacoustic music practitioners. We identified these group of users as potential early adopters of the MM due to their familiarity with electronic and digitally-mediated musical practices. This article can be considered as a value discovery involving design fiction and community-based design methodologies. We considered the MM as both a musical artifact and a platform that needs to be defined and specified. To better explore these aspects, we conducted three workshops.

The first workshop involved the faculty and students from the electroacoustic music curriculum of the Conservatory 'A. Casella' of L'Aquila (Italy). The results of this first workshop have been analysed and presented in a dedicated publication (Boem et al. 2024). However, while involving potential users in an early phase of research can enable researchers to understand the opportunities and challenges of new technologies, this approach has shown some criticalities, such as users not always being aware of their own needs and desires (Hussein, Mahmud, and Tap 2014; Strömberg et al. 2018). To mitigate this issue, we included two other groups of future early adopters and stakeholders of the MM: classically trained musicians, and producers and sound engineers. These two groups came respectively from the curriculum of instrumental music from the Conservatory 'A. Casella' of L'Aquila (Italy), and from the curriculum of sound engineering from the SAE Institute Milano (Italy).

By examining the Metaverse through the lens of musical technologies, we aim to offer valuable insights for both academic and industry professionals. Understanding this perspective is essential for identifying the requirements that can enable authentic real-time interactions and support rich human expression within the Metaverse. Musical applications, by their very nature, demand strict coordination, precise timing, and a balance between technical accuracy and expressive nuance. These requirements make them a valuable testbed for addressing broader challenges in Metaverse development. Moreover, musical applications, by their very nature, demand strict requirements such as coordination, precise timing, and a balance between precision and expression. Therefore, these requirements make musical applications a valuable testbed for tackling challenges and issues related to the development of the Metaverse as whole. Following the call of Morreale et al. (2020), our work explores the idea of a Metaverse for musical applications by involving future early adopters and stakeholders to understand and analyse their perceptions and values. This approach aims to avoid the techno-solutionism that has shaped much of the research on the Metaverse. Our goal is to develop a deeper reflection within the music technology community on this topic and imagine how to create a Metaverse better suited for people interested in composing, playing, and learning music in the twenty-first century.

## 2. Materials and methods

We conducted three half-day workshops (Ørngreen and Levinsen 2017) centred on group activities designed to elicit participants' needs through personal and group reflection processes. The workshops were held at two Italian institutions: Conservatory 'A. Casella' of L'Aquila and the SAE Institute Milano, involving three distinct participant groups. During each workshop, participants reflected on three dimensions of their musical practice:

• **Composition**: a set of activities and practices which encompasses processes related to creating and composing musical pieces. • **Performance**: a dimension of music-making that includes aspects such as executing pieces before an audience and the associated preparation activities (e.g. rehearsals). • **Education**: the practices and activities related to music education, including music theory, music history, and pedagogy.

### 2.1. Theoretical framework and dimension selection

The three dimensions examined in this study (Composition, Performance, and Education) were selected based on converging evidence from multiple established frameworks in music technology and XR research.

A 2021 survey has identified several primary functions of Musical Extended Realities (Turchet, Hamilton, and Çamci 2021) including Performance, Education, Entertainment, Development, Perception Study, Composition, and Sound Engineering. These correspond to target user groups such as Performers, Students, Audience members, Developers, Composers, and Sound Engineers.

Since our study targeted music practitioners (composers and performers) within an educational context, we selected the three dimensions that directly align with our participants' professional activities: Performance, Education, and Composition. Moreover, these dimensions represent the core areas where our participant groups engage professionally and where immersive technologies such as AR/MR/VR can provide a positive impact (Gómez-Sirvent et al. 2024; Serafin et al. 2016; Turchet, Hamilton, and Çamci 2021).

### 2.2. Workshop design and implementation

Since the MM refers to a vision not yet fully realized, we asked participants to reflect on each dimension 'as if by magic' the MM already existed. Drawing on previous work from the communities of Human–Computer Interaction and New Interfaces for Musical Expression (NIME) (Andersen and Wakkary 2019; Lepri and McPherson 2019), we used the term 'magic' as a device for eliciting subjective experiences from participants, encouraging them to think beyond the limitations of current technologies.

To facilitate the formulation and exploration of their ideas, we provided each participant with a piece of A4 paper, which we referred to as a Sketch Sheet, on which participants were instructed to write and draw their reflections.

### 2.3. Participants

Participants were recruited through open calls at two different institutions. All participants were native italian speakers. The first two workshops were conducted at the Conservatory of L'Aquila. The first workshop was attended by a total of 14 participants (1 female, 13 males; mean age = 38, SD = 11.22). Participants included faculty, current, and former students with Bachelor's and Master's degrees in 'Electronic Music and New Musical Technologies.' The second workshop was attended by a total of 15 participants (8 females, 7 males; mean age = 35.6, SD = 16.89). All participants were current and former students with Bachelor's and Master's degrees and were classical musicians. The third workshop was held at the SAE Institute, involving a total of 14 participants (3 females, 11 males; mean age = 30.71, SD = 7.02). Among all participants, five reported previous exposure to XR technologies (two in the first group, and three in the third group). However, their experience was sporadic and limited to one or two trial sessions mainly with videogames. These demographic data were assessed through a paper-based survey that was distributed to participants.

### 2.4. Procedure

Each workshop followed a standardized protocol with defined facilitator roles. The research team consisted of three members: two facilitators (primary and secondary) and one note-taker. The primary facilitator led dimension introductions and guided the discussions, while the secondary facilitator monitored group dynamics and provided clarifying questions. The note-taker documented verbal responses of the participants.

The workshops lasted a total of 4 hours each, including a 15-minute break. At the outset, participants were asked to sign a consent form after receiving a comprehensive explanation of the workshop's purposes. The workshops began with a 30-minute seminar introducing the MM domain, presenting foundational concepts including the Reality-Virtuality Continuum (Milgram and Kishino 1994), Spatial Computing (Shekhar, Feiner, and Aref 2015), and immersive audio (Paterson and Lee 2021). Then facilitators followed this sequence: at first, the they provided a brief explanation of the three dimensions examined during the workshop within the context of the MM. Then, the primary facilitator provided a verbatim reading of a prompt: 'How do you see [dimension] in the MM if "by magic" it was available now?' This was also presented visually to the participants, through video projection. The prompt was adapted for each dimension: for Composition the prompt reads like this 'How do you see your compositional practice in the MM if "by magic" it was available now?', for Performance it became 'How do you see your performance practice in the MM if "by magic" it was available now?', for Education 'How do you see your approach to music learning and or teaching in the MM if "by magic" it was available now?'.

Then, participants were given 10 minutes to note their ideas on the Sketch Sheets. This was followed by a presentation phase, were one at a time, participants were asked to freely discuss their ideas and reflections with the others. All workshops were conducted in the Italian language. In order to maintain neutrality, throughout the duration of each workshop the primary and secondary facilitators avoided leading the discussion, instead focusing on reframing concepts and ideas proposed by participants rather than suggesting new ones. Interventions were limited to clarification requests and process management.

Our research complies with the principles of the Declaration of Helsinki, and the Ethical Standards of the NIME Conference (Morreale et al. 2023). Since the risk level concerning data treatment was considered low by our institution, we did not require approval from the Ethics Committee.

**2.5. Data collection and analysis**

All group activities were recorded using audio and video equipment. The facilitators also collected the Sketch Sheets. Along with the notes taken during each workshop, a reflexive thematic analysis (Braun and Clarke 2019) was performed on the collected data based on grounded theory (Glaser and Strauss 2017).

Prior to analysis, audio recordings were transcribed and participant-generated texts were digitized. These data, together with researcher notes, were stored in a shared Google Drive folder.

**2.6. Themes identification**

The process of theme identification involved a structured consensus process among three of the authors of the paper.

First, each author independently conducted open coding on all transcripts, notes, and sketch sheets using thematic analysis principles. Initial codes were descriptive labels that aimed at capturing discrete ideas and concepts mentioned by participants. Examples of initial codes were 'spatial sound placement' 'creation tools' (for Composition), 'virtual control surfaces' 'audience perception' (for Performance), 'remote access to lessons' 'visualization aids of concepts' (for Education). Authors worked separately to ensure diverse analytical perspectives.

Second, the three authors conducted three in-person meetings. Each meeting addressed one dimension for each workshop. The process was repeated two times, in different days. During these meetings, the authors presented their independent findings, discussed convergences and divergences, and iteratively refined the proposed thematic categories and their boundaries. Subsequently, initial codes were grouped to form broader categories through iterative analysis. For instances, codes such 'monitoring and understanding physical expression', 'self evaluation', 'testing performative practices of contemporary music' were grouped under the theme 'Interactive Practice and Performance Optimization' for the Performance dimension of Workshop 2.

Third, a final meeting was held to resolve the remaining disagreements. When they occurred, authors re-examined the source material together. Final themes were selected when they appeared to be: prevalent across multiple participants (at least three) within each group, formed a coherent conceptual category, and were distinct from other themes. All three authors had to agree on final theme definitions and boundaries.

Finally, for this publication, we translated all keywords and quotes into English using a combination of Google Translate and DeepL Translator.

## 3. Results

We present the results of the thematic analysis first for each group of users. Finally, we discuss common themes across groups.

### 3.1. Workshop 1: electroacoustic composers

#### 3.1.1. Composition

For this dimension, we have identified three essential themes that relate to how participants framed their creative processes in the context of the MM.

• **Space as a compositional dimension.** Eight participants (P3, P4, P6, P7, P9, P10, P13, P14) highlighted the spatial dimension as a central element of the MM. They specifically noted the potential of virtual spaces to serve as non-linear, interactive compositional environments (e.g. P4: 'where users can actually enter inside the work, stay inside the visuals, and inside the sound'). This suggestion was further expanded by P7: 'the possibility of simulating an acoustic space in which I can move and where there are nodes, and there are points where there are sound sources from which I can approach and/or move away.' Furthermore, the participants discussed how immersive and virtual spaces could be utilized to create audio-visual installations (e.g. P9: 'Real applications are installations, not composition').

• **Virtual interfaces.** Seven participants (P5, P6, P7, P10, P11, P13, P14) envisioned the MM as capable of offering novel virtual interfaces for compositional and sound processing, which should be radically different from those available in the physical reality and unique to the virtual world, as suggested by P6: 'I imagine it as a medium/interface where I can use tools that I don't have available but which I can virtually obtain'. Such interfaces should be used not only for rapid prototyping of virtual instruments but also for testing compositional structures and tools for annotation (e.g. P11: 'Annotation tools could be useful in a performance environment, where we are used to taking notes while performing'). Moreover, participants emphasized the need for these interfaces to be customizable and flexible. This would enable users to tailor their creative environments to suit better their artistic needs and preferences (e.g. P7: 'I would like to have the possibility that this Metaverse is customizable and consistent with my way of creating').

• **An immature media space.** According to three participants (P7, P11, P13), an understanding of the technical and conceptual aspects of the MM should precede any attempt to evaluate its impact on compositional practices. P11 raised a pertinent question 'What is the utility we can find in controlling, from an XR environment, a real environment for music production?'. For these participants, such kind of inquiry reflects the current state of the MM, which P7 describes as 'a field of total experimentation', but it is also perceived as far from being a mature and defined medium for compositional practice. This signals a cautious approach to integrating the MM in processes related to music composition.

### 3.1.2. Performance

In the domain of performance, four central themes emerged from our analysis:

• **Multisensory enhancements.** Four participants (P1, P4, P6, P10) emphasized the importance of multisensory experiences in the context of performance. P10 commented: 'The goal is: the performer not only has control over a musical parameter but over multisensory parameters'. According to participants, integrating visual, auditory, and tactile stimuli could provide musicians with vital information about their physical and emotional state, and the auditory components of a live performance, which are typically the main focus of both composers and performers. (e.g. P6: 'I imagine something that can support that part of the performance that cannot be communicated through sound or body movement like the performer's emotions, and therefore, thanks to this, the listener's thoughts on what I am communicating in the act is also clarified performative').

• **Performance preparation in the physical-virtual space.** Four participants (P1, P5, P10, P13) recognized the significance of allowing performers to use Spatial Computing technologies to arrange virtual instruments, controllers, and parameters in both VR and MR environments to meet their specific needs in the preparation of a concert (e.g. P1: 'I imagine myself having greater control over performance practice, thanks to spatial computing and therefore the possibility of placing the controls I want wherever I want in space'). Furthermore, this approach includes the positioning and arrangement of sound sources in MM environments (e.g. P13: 'I would like to place my virtual loudspeakers based on how the room is displayed').

• **XR as a tool for behavioural research.** An element that emerged from four participants (P1, P9, P10, P13) was the use of XR as a sensing technology that can monitor the physical state, posture, and expressivity of a performer, thereby enhancing and extending their capabilities (e.g. P9: 'achieving an augmented performer'; P1: 'study and in-depth analysis of posture and movements'; P10 described this as a form of 'performance ethnography').

• **Performers' Skepticism.** During the discussion phase, several participants questioned the validity and utility of the Metaverse for musical performance. Six participants (P3, P7, P8, P12, P13) expressed doubts about the MM's ability to realistically substitute for physical performance, at least in a VR-only setting (e.g. P9: 'I can't imagine a performance practice in the Metaverse capable of replacing the physicality, or direct contact with an instrument, whatever it is… '). Furthermore, P12 suggested that this dimension should be better experienced using AR since it 'guarantees a co-presence of real bodies and therefore guarantees this feedback'. This perspective highlights a preference of composers for AR —and MR as well — environments over completely virtual ones, emphasizing the importance of physical presence in performance settings.

### 3.1.3. Education

Within this dimension, we identified four main themes:

- **Immersive learning.** Most participants (P1, P2, P3, P4, P5, P6, P7, P9, P10, P11, P12, P13, P14) articulated their vision of education within the MM in terms of 'immersive learning'. Both P1 and P9 highlighted that the MM could facilitate a remote learning environment that is 'more immersive and multisensory, without all the problems that classic 2D software such as Zoom introduces' (P1). Furthermore, P6 and P7 envisioned the MM as a space where immersive audiovisual recordings of classes could be accessed, revisited, and re-experienced later. Because of its social dimension, two participants (P10, P13) pointed out that the MM seems particularly adapt to teaching group improvisation.

- **Learning simulators.** Eight participants (P1, P2, P5, P6, P7, P10, P11, P12) saw the MM as a space suited for developing 'virtual classrooms' (P5) tailored to specific activities and learning materials. For example, these classrooms could enable students —either alone or with teachers— to analyse and rehearse virtual versions of electroacoustic compositions. (e.g. P11: 'I would like to build a multimodal environment where together we can comment, analyse, and better understand a particular piece or performance'). Additionally, the MM could facilitate the study of specific topics such as acoustics. Through virtual replicas of existing physical spaces (e.g. auditoriums and concert halls), students could experiment with the relationships between sound and space in more controlled settings. Similarly, five participants (P3, P4, P9, P11, P12) recognized the potential of the MM for studying the history of electroacoustic music. They suggested that the MM could be used to replicate studios that no longer exist or to model historical instruments that are difficult or costly to replicate in real life 'as a sort of simulator' (P9). As expressed by P12: 'a VR environment that allows me to visit places, decayed or no longer existing places that allow me to review/retrace historical moments'. This may be particularly useful with spatial-based historical works of electroacoustic music, such as Luigi Nono's 'Prometheus' (P4). Participants felt that immersive experiences, which integrate learning about the history of electroacoustic music, were more effective compared to more accepted means like listening to audio recordings or watching video documentation of specific pieces.

- **Perceptual aids.** Five participants (P4, P9, P12, P13, P14) suggested psychoacoustics as a subject that could be effectively studied in the MM. Participants considered the study of psychoacoustics was considered a critical aspect of electroacoustic music education (e.g. P4: 'It would be much more practical to understand acoustic phenomena […] such as diffraction or reflection'). According to participants, in MM environments, acoustic phenomena could be experienced through the use of real-time 3D visualizations, which serve as perceptual aids that could facilitate the learning of complex theoretical processes (e.g. P14: 'Visualize theoretical processes that require a level of abstraction and which perhaps are not so direct to learn in the early stages. I think that through a 3D visualization in a multiuser environment, these concepts can be learned more easily').

• **Inclusiveness.** Five participants (P1, P7, P9, P10, P11) emphasized the necessity of designing the MM as inclusive as possible, particularly for teaching and learning purposes. Firstly, they proposed that the MM should primarily serve as a social space, where 'one can do research or talk about music [...] or meet new people, which is a classic thing that always happens in typical conservatories' (P1). Secondly, the MM could provide off-site students access to virtual replicas or 'digital twins' of their music institution, which 'would allow for great economic advantages' (P7). Thirdly, participants stressed the importance that a MM should support and foster 'critical thinking' (P10), which was seen as a fundamental component of musical education. However, there was a concern that a heavily technology-mediated environment like the MM could exacerbate the digital divide between students and institutions (P10: 'We should think about low-cost and easy-to-use hardware/software systems […] because this could lead to a big danger that could result in a social division between those who can access these technologies and those that cannot').

### 3.2. Workshop 2: classical musicians

#### 3.2.1. Composition

For this dimension, we identified one theme that highlights how participants framed their compositional practices in relation to the MM.

• **Interaction Between Composer and Performer:** Four participants (P3, P9, P10, P14) reflected on the possibility of the composer-performer relationship in the context of immersive environments. In the MM, the composer could interact directly with a virtual performer, asking for advice on musical gestures or exploring how musicians physically produce sound (P9: 'In the Metaverse, a composer could ask the virtual performer something like, "If I want a heavy gesture on this note, what would you suggest?" This opens up new possibilities for dynamic interactions. It would also be fascinating for composers to understand what happens inside a musician's body during a performance, such as what goes on inside a flutist's mouth when producing a sound or how a cellist executes a particular gesture'). This would preserve the traditional dynamic while also introducing new possibilities for real-time, adaptive collaboration in a virtual environment.

#### 3.2.2. Performance

For this dimension, we identified two key themes that capture how participants framed their performative practices in relation to the MM.

• **Interactive Practice and Performance Optimization:** Six participants (P5, P6, P9, P14, P15, P5) considered the use of immersive and interactive technologies for self-improvement and performance enhancement in their practice. They envisioned scenarios where musicians can practice with their own avatar for self-assessment (P6: 'Trying to play with an avatar that plays exactly what I am playing, and then the moment it plays what I am playing, I stop playing, the avatar continues, and so I do a kind of self-assessment'), or by using real-time sonifications to guide and improve physical movements (P15: 'The possibility of sonorizing the movements (through sound) to determine what movement to do and vice versa, and then sound could help improve performance postures'). This suggests a desire to use technology as a tool for both practice and performance optimization, particularly in addressing common challenges like posture (P9: 'It could help in performance to focus on the attack of the violin's shoulder, like in the positions since we are the backward type, we do not see well or even the director').

• **Acoustic Environment Control:** Four participants (P9, P10, P15, P17) expressed interest in having virtual spaces where they could experience different types of acoustics. This aspect was considered for two elements. The first is a way to facilitate the listening experience for both the performer but also for the audience to create a situation that can better deliver the auditory part of the concert without the problems that sometimes can emerge in live situations (P9: 'another problematic aspect for the orchestra is the acoustics; therefore, the Metaverse could adapt the sound to reach the listener better'). The second is the use of virtual acoustics to recreate the conditions of places such as concert halls or theatres that don't exist anymore or that are difficult to reach, maybe through a virtual reconstruction (P15: 'Reconstruct sound spaces that are no longer there, like past performances, places inaccessible today, etc… ').

• **Enhanced Virtual Rehearsal Spaces:** Three participants (P6, P8, P12) focused on the aspect of the performance's preparation and imagined that virtual environments could facilitate the rehearsal part by simulating an experience or creating a more immersive experience with other musicians in different locations: 'I could rehearse at home, but I actually have a stage, an orchestra, and a director, and it might stimulate me to get into the role/part because at home individual practice gets boring after a while' (P12). This can also include the preparation of particular pieces and techniques, especially the ones used in contemporary music (P6: 'Create a rather "smart" environment for a performer who wants to approach contemporary music (e.g. prepared piano) and do so in an intelligent (smart) environment and thus with extreme ease on the part of the performer').

• **Performance Anxiety Management:** Three participants (P10, P16, P17) highlighted immersive environments as a tool for managing and overcoming performance anxiety in preparation for a performance, which is a condition especially experienced by students but to a certain extent by musicians of different expertise. This can be done by 'rehearsing at home as if you were in front of a large audience' (P16). This suggests musicians see potential in virtual practice environments as a safe space to build confidence and reduce stage fright before real performances gradually.

### 3.2.3. Education

For this dimension, we identified two key themes that illustrate how participants framed their ideas about the educational dimension in the context of the MM.

• **Virtual Teaching Support:** Four participants (P7, P15, P1, P6) expressed the desire to have a personalized music education system that can simulate the presence and guidance of a human teacher. Participants envisioned that the MM could provide them immediate, interactive feedback similar to having a teacher present (P1: 'In the study phase having his/her [the teacher] image in front of me repeating those 2–3 key words can be essential'). This includes accessing recorded teacher instructions, as well as receiving real-time corrections on technique, and getting visual demonstrations of musical concepts in three dimensions (P6: 'As a violinist, on the other hand, in the study phase, it would also be helpful to have an instrument that could correct the interaction that it is as if you have the teacher in front of you telling you what you are doing wrong').

• **Learning simulators:** Four participants (P2, P3, P12, P4) expressed the possibility of the MM to democratize and expand access to musical education and instruments. Participants envisioned the use of immersive technologies to overcome the high costs and accessibility of particular musical instruments and to gain unique learning opportunities, such as visualizing internal physical processes like the mechanics of the vocal tract for singers (P12: 'Imagine and think that an organ costs 150000 euros; imagine you can play it in the Metaverse with the ancient temperament for only 1000 euros'). This theme emphasizes how virtual technology could make previously inaccessible or expensive musical experiences more affordable and provide innovative ways to understand complex musical concepts.

### 3.3. Workshop 3: producers and sound engineers

#### 3.3.1. Composition

For this dimension, we have identified two themes that relate to how participants framed their compositional processes in the context of the MM. From this point forward, we will use the term 'producers' to collectively refer to both producers and sound engineers.

• **Immersive Studio Production:** An element that emerged from eight participants (P2, P3, P5, P8, P10, P12, P13, P14) centred on the desire for an immersive and user-centred virtual music production environment. Producers envisioned a workspace where they could be physically at the centre of their creative process and surrounded by accessible virtual tools and instruments rather than being limited to traditional computer-based interfaces (P2: 'The PC is at the center of the production process … I would like to move to the center instead. I would like the DAW around me and me at the center […] so all the tools around me in order to have more immersive audio feedback'). For participants, such environments should also allow collaboration and sharing with other producers and musicians: 'I would like to have templates of different studios where I can work on projects with other colleagues and having some rooms where other people can access them at certain times' (P3).

• **Multisensory Songwriting:** For six participants (P1, P4, P6, P10, P11, P12), a notable aspect of the MM for enhancing the compositional process is the ability to create customizable and inspiring virtual environments for music production and collaboration. Participants envisioned spaces that can go beyond traditional studio settings, wanting the freedom to create in diverse, mood-enhancing virtual locations while maintaining access to all necessary tools and instruments (P11: 'Total immersion in the environment-story, like having references; I would like to write music where I do not just write, but I find myself in that environment that I am writing and therefore touch what I am writing with all my sensory dimensions. For example, s/he talks about Japanese flowers in my song, and I would like to have an environment that makes me smell the flowers').

### 3.3.2. Performance

For this dimension, we identified two key themes that reflect how participants conceptualized their performative practices within the framework of the MM.

• **Fluid Performance Spaces:** Nine participants (P1, P4, P5, P7, P9, P11, P12, P13, P14) reflected on how live music experiences will be transformed through enhanced immersive technologies. They envisioned interactive performances where music dynamically shapes the visual environment, creating immersive, multi-sensory experiences that extend the traditional concert experience. For participants, the focus is on allowing audiences (being physically present or connecting remotely) to experience personalized atmospheres that complement and enhance the musical performance in the MM (P1: 'An artist in the center with the whole audience around; a kind of open field very flat and without music, and then after an artist creates his music from scratch, the whole world around is modulated, and dimensional visuals are created that give an extra aspect to the music').

• **Audience Response:** Seven participants (P1, P2, P3, P4, P5, P7, P8) emphasized the desire for enhanced performer-audience interaction in immersive performances. These participants would like to have technologies that can enable them (as musicians) to gauge and respond to audience reactions in real-time, similar to traditional concerts but with enhanced capabilities (P1: 'The Metaverse must give the possibility to interact with the audience, perhaps by modifying the environment or promoting events or happenings'). They envision a dynamic environment where performers can more easily understand audience engagement, modify the performance space based on crowd response, and then receive meaningful feedback about audience reactions (P8: 'I would like to have a sort of feedback from the audience -e.g. perception of what the audience is doing- it is already done now with ambient microphones to understand how happy or not the audience is'). This suggests a desire to maintain and potentially enhance a two-way connection between artist and audience that is considered crucial.

• **The Virtual Concert Experience Gap:** Three participants (P1, P3, P10) expressed their concerns about the authenticity and limitations of performances in the MM. They expressed apprehension that virtual environments might compromise essential elements of live music experiences (P3: 'I am wary of the concert in the metaverse because, in my opinion, many elements are lost. I think it becomes more of a visual show than a listening one (e.g. in a concert, there is a particular sound system, a particular space, etc.)'). They also worry about the heavy association of the Metaverse with gaming that can then potentially trivialize the serious, authentic nature of live musical performances, suggesting a fundamental skepticism about whether virtual spaces can truly replicate the depth and richness of real-world concert experiences.

### 3.3.3. Education

Regarding this dimension, we identified two main themes:

• **Virtual Audio Equipment Training:** For nine participants (P1, P3, P4, P5, P7, P8, P9, P13, P14), an important element for the MM lies in the possibility of using virtual simulations of audio equipment and experimenting with recording techniques and tools. These participants described the use of immersive environments for learning and practicing with expensive or hard-to-access audio hardware and gaining hands-on experience with studio equipment (P3: 'You could instead use it to do simulations (e.g. acoustics, studio simulation, the signal flow of that specific studio, etc.). How to do the signal flow of specific hardware? I imagine an environment that helps me understand the signal flow of particular hardware without actually having them but having a digital twin of the virtual ones and seeing their simulation'). This theme emphasizes interactive learning through virtual tools as an alternative to the use of technical manuals and as a way to overcome limited physical access to professional equipment, allowing for risk-free experimentation and practical understanding of audio engineering concepts (P14: 'Instead of a user manual, you have a link that gives you the digital twin of the equipment and the workflow on XR; you learn how it works in a 3D/immersive way, and obviously, you do not have the natural feeling of the equipment, but it gives you a much more immersive and practical idea than reading hours and hours of user manuals').

• **XR masterclass:** Three participants (P4, P6, P12) envisioned interactive sessions where they could virtually attend masterclasses, observe professional workflows in real studios, and even practice or perform with digital recreations of legendary artists. For them, social immersive environments could democratize access to expert knowledge and mentorship, allowing students to learn directly from industry leaders regardless of geographical limitations (P6: 'Do masterclasses of the best US sound engineers without making them travel. They are done in XR in the metaverse in his/her Studio, and you see how s/he moves from the beginning to the end of the project, learning workflow, moving inside their studio').

## 3.4. Common themes

After identifying the main themes for each dimension of the MM (Composition, Performance, and Education), we examined the convergence of themes across the different groups of participants. Our analysis reveals distinct patterns in how they envision the MM. The distribution of responses across these dimensions shows notable variations among Electroacoustic Composers, Classical Musicians, and Producers. Within the Composition dimension, Classical Musicians engaged less than other groups, and we found only one consistent theme. The Performance dimension revealed more balanced participation, with Electroacoustic Composers showing the highest clustering of themes, followed by Classical Musicians and Producers. In the Education dimension, we found consistent participation and clustering across all groups.

We then further analysed each dimension to identify common themes among groups of participants.

### 3.4.1. Composition themes

In the Composition dimension, each group approached the MM with distinct perspectives. However, we noticed that the collaborative and social aspect of composition that the MM could facilitate, is still not well considered by all groups, and it did not became a prominent or unique factor. The analysis reveals two significant cross-group themes:

**Composer-Performer Interaction:** The different groups focus not on the act of composition (the interaction between a composer and the musical material they are organizing) but on the composer-performer interactions. Each group brings unique perspectives to this theme. Classical Musicians envision direct interaction with virtual performers, enabling immediate feedback about specific gestures and physical aspects of sound production. Electroacoustic Composers focus on the development of virtual interfaces to facilitate rapid prototyping and testing of new compositional structures. Producers emphasize the potential for collaborative virtual spaces where multiple creators can work together in real-time.

**Spatial Creation:** This emerged as a creative dimension across all groups. Electroacoustic composers particularly emphasize virtual spaces as non-linear and interactive compositional environments where users can 'enter inside the work' and move within the sound space, allowing for unprecedented interactions with sound sources. They look at the MM as a possibility for exploring new dimensions and possibilities that appear to them more related to the practice of sound art, such as installations, diverging from a more linear approach to composition. Producers seek to transform traditional computer-centred production into a more human-centred environment, envisioning themselves physically at the centre of their creative process and surrounded by virtual tools and instruments rather than confined to conventional computer interfaces (e.g., flat screens).

### 3.4.2. Performance themes

The Performance dimension reveals varying priorities among the groups while maintaining some common ground. Three major themes emerge across groups:

**Skepticism about Virtual Performances:** This appears to be the prominent theme, particularly among Producers and Electroacoustic Composers, who express concern about potentially losing essential elements of live music experiences, such as physicality and the sense of shared experience.

**Virtual Acoustic Control:** This emerges as a crucial theme across all groups. Classical Musicians seek to experience different types of acoustics of environments and recreate historical venues, while Producers envision dynamic performance spaces that can be adjusted in real-time. Electroacoustic Composers focus on arranging virtual sound sources in virtual environments and controlling their spatial parameters.

**Enhanced Feedback Systems:** This theme manifests differently for each group. Producers seek real-time monitoring of audience responses and novel interactions. Electroacoustic Composers emphasize developing multisensory feedback specifically tailored for performers. Classical Musicians focus on feedback mechanisms oriented toward optimizing performance and enabling techniques for self-assessment.

### 3.4.3. Education themes

**Learning Simulators:** In the Education dimension, all groups share a fundamental interest in themes regarding learning simulators, though they approach this potential from different angles. Electroacoustic Composers emphasize the creation of multisensory environments that can help them with understanding acoustic phenomena through three-dimensional and real-time simulations. They also focus on inclusivity, aiming to broaden access to musical education on topics such as 3D audio and perception. Producers take a more technically driven approach, emphasizing interactive masterclasses with experts and practical demonstrations of workflow for music production through digital twins. Classical Musicians concentrate on teacher-student relationships, indicating a stronger need for supporting interpersonal interaction compared to the more self-sufficient and technical approaches favoured by Producers and Composers.

**Democratization of Access:** This theme appears consistently across all groups but with varying emphases. Electroacoustic Composers stress inclusive design and access to virtual replicas of music institutions. Classical Musicians focus on accessing expensive instruments and performance spaces. Producers highlight the potential for widespread access to professional studio equipment and expert knowledge.

### 3.4.4. Theme relationships and overall findings

Several themes interrelate and reinforce each other across different dimensions. The overall skepticism about virtual performance often connects to concerns about acoustic environment control. Enhanced feedback systems support both performance and educational goals. The democratization of access influences all aspects of musical activity in the Metaverse.

Our analysis reveals that while each group maintains distinct priorities based on their musical roles, significant overlap exists in their fundamental concerns and aspirations for the technology. All groups view the Metaverse as a tool for enhancing rather than replacing traditional musical practices, with particular emphasis on practical applications in education and preparation rather than final performance delivery.

A consistent preference for Augmented Reality and Mixed Reality over purely virtual environments emerges across all groups, suggesting that maintaining a connection to physical reality remains crucial in musical activities. Furthermore, each group's professional background significantly influences their approach across all dimensions: Electroacoustic Composers apply their experimental and spatial thinking, Classical Musicians consistently emphasize physical connection and relations to instrumental techniques, and Producers focus on workflow and accessibility across all aspects of the technology.

Most significantly, all groups identify the Metaverse's greatest potential in preparatory and educational contexts rather than as a replacement for traditional performance spaces. This suggests that for these potential users, the Musical Metaverse might have its most profound impact on how musicians learn and prepare for their performances rather than on how they ultimately present their work.

## 4. Discussion

This study reveals several key insights on how musical practitioners envision and understand the potential of the MM. The findings suggest that while there is significant enthusiasm for certain applications, particularly in the context of education and performance preparation, there are also important concerns and limitations that need to be addressed for the MM to become a viable platform for musical creativity and expression. Moreover, the results align with the observations we found in the previous analysis of the Electroacoustic Composers group alone (Boem et al. 2024).

### 4.1. Educational applications as primary use case

The strongest consensus among participants emerged around the educational potential of the MM. The concept of 'learning simulators' appeared consistently across all groups, though with different emphases reflecting their specific domains.

These results resonate with the principles of 'Immersive Learning' already explored within XR (Makransky and Petersen 2021; Petersen, Petkakis, and Makransky 2022), where immersive technologies serve as enablers and amplifiers of educational experiences. Although music education has already adopted immersive technologies, their implementation has typically been confined to single-user scenarios and has not considered the social and multi-user types of interactions that the MM can afford (Johnson, Damian, and Tzanetakis 2020; Orman, Price, and Russell 2017).

This suggests that the initial development of the MM might be most productively focused on educational applications, where the benefits of virtualization (cost reduction, accessibility, repeatability) appear to outweigh the drawbacks of reduced physical presence. The findings indicate three primary educational applications: historical recreation and experiential learning, enhanced perceptualization of complex musical concepts, and technical skill development through virtual simulation.

#### 4.1.1. Social and pedagogical framework

The importance of education, among other dimensions, might be understood through Lefebvre's concept of social space: the 'classroom' represents a uniquely social environment shaped by the dynamic relationships between students and professors. This social character aligns naturally with the defining elements of both traditional education and the Metaverse itself. While the concept of 'learning simulators' appeared consistently across all groups. Participants also emphasized that the MM should support and foster critical reflection rather than mere simulation.

#### 4.1.2. Evidence from VR/AR research

Previous research on shared VR environments has demonstrated their effectiveness for developing general research and inquiry skills, as well as experimenting with high-level concepts and processes. Especially, studies have shown improved motivation and sociability among distance learners in shared virtual worlds (Zhang 2024). While these findings mirror existing proposals for VR and AR applications (Nijs and Behzadaval 2024; Serafin et al. 2017), our results suggest that musical education deserves more focused exploration to develop methodologies suited to specific musical curricula.

#### 4.1.3. Key educational applications

The findings indicate three primary educational applications:

**Historical Recreation:** this theme includes immersive exploration not only of music history, encompassing both analytical study of compositions and investigation of material and technical culture.

Participants offered concrete implementations, particularly in the context of electroacoustic music. They proposed creating virtual replicas of significant 'technological environments' (Di Scipio 2000) that shaped the history of electroacoustic music, such as the 'Philips Pavillon' by Le Corbusier and Xenakis or the 'Studio di Fonologia' of Milano. These replica of historical spaces would serve not just as study environments, but as creative spaces for exploration and analysis, leveraging the spatial and multimodal characteristics of the MM. Such environments would allow students to engage with both the theoretical and practical aspects of musical heritage while developing their analytical and creative skills in an immersive context. While similar ideas have already been explored in the field of Musical XR (Lombardo et al. 2009), the social, collaborative, and persistent nature of the MM has not been considered.

**Perceptualization of Musical Concepts:** A particularly promising application emerged around what we defined as 'data perceptualization', a concept originating from biomedical studies (Lindborg, Chopra, and Groß-Vogt 2023). While this approach extends beyond education, participants highlighted its particular value in educational contexts for understanding topics such as the properties of sound propagation in space. Unlike traditional spectral and physical tools for sound analysis, immersive technologies enable three-dimensional exploration of sound in its components. Previous research has explored such perceptualizations using AR, VR, and MR for visualizing sound fields in an immersive manner (Arslan et al. 2022; Deines et al. 2007; Inoue et al. 2019). In the MM context, these perceptualizations can serve a dual purpose: supporting theoretical explanations of acoustics while enabling students and teachers to test hypotheses by modifying simulation parameters.

**Skill Development:** Another application envisioned is the use of immersive environments for practical learning and the development of musical skills. While this appears to be an essential aspect of music education, it is not clear which kinds of skills can be practically developed in the context of the MM. Within current VR applications such as PatchWorld, it is possible to attend group classes on the use of custom virtual instruments. In TribeXR, it is possible to learn DJing and mixing techniques. MR was explored for helping with piano skill learning with a co-present and superimposed remote teacher (Gerry, Dahl, and Serafin 2019). However, despite these examples, skill development represents an open topic that deserves careful consideration, mainly when it should be used in a musical curriculum. Should it be used for learning basic and general skills such as tempo and rhythm, or should it be applied to specific instruments? Is it more effective for novices or for experts? This aspect requires further research.

### 4.2. Democratic access vs. authentic experience

A central tension emerged between the MM's potential to democratize access to musical resources and the concern about maintaining authentic musical experiences. All three groups recognized the transformative potential of the MM to provide access to expensive instruments, rare historical artifacts, and experts' knowledge. The concept of inclusiveness emerged as a shared value among participants, highlighting the hope of making learning accessible to students regardless of their economic situation and geographical location.

However, this potential for democratization was consistently balanced against two major concerns. First, participants expressed skepticism about the authenticity and quality of virtual musical experiences, particularly in the context of performance, where the MM's ability to replicate the nuanced qualities of live musical interaction was questioned. Second, participants raised concerns about potential barriers that might be introduced by an educational system that becomes heavily dependent on ubiquitous technologies. Here, inclusiveness was considered a value to protect, counteracting potential dangers such as a possible creation of a digital divide.

These concerns were informed by recent experiences during the COVID-19 pandemic, which revealed how remote and technology-mediated learning can introduce inequalities in the educational process across several parts of the world (Azubuike, Adegboye, and Quadri 2021; Vigevano and Mattei 2025). In the current socio-political landscape, this tension between democratization and potential barriers created by new technologies represents a critical aspect that must be considered when designing a 'Good Metaverse' (Zallio and Clarkson 2022). These considerations align with recent discussions on the ethical dimensions of the MM (Young 2024) and call for further examination in the context of music technology education and social immersive platforms more broadly.

### 4.3. Spatial computing as a new creative dimension

The integration of spatial computing within the MM marks a transformative moment in the evolution of musical practice, representing more than just a mere technological advancement. While musicians have traditionally worked with fundamental elements of sound such as frequency, dynamics, and tempo, our analysis reveals that space itself emerges as a new and primary compositional parameter, fundamentally altering how we conceive, create, and experience music.

This transformation builds upon a rich historical foundation in electroacoustic music, where composers have long explored the concept of 'composing with space' (Basanta 2015; Di Scipio 2011; Lopes and Guedes 2020; Macedo 2015). These explorations have traditionally transcended simple spatialization techniques, delving into the complex relationships between sound and physical environments. Our research, drawing from in-depth conversations with composers and producers, reveals how spatial thinking has become deeply integrated into contemporary creative processes, with practitioners viewing spatial arrangements as fundamental compositional elements rather than secondary consideration.

The pioneering work of several composers has demonstrated the potential of collaborative virtual space as both a scoring systems and performance instruments (Bell 2023; Ciciliani 2020; Dziwis and von Coler 2023; Hamilton 2019; Martín 2018; Pirchner 2020). However, our analysis suggests that the MM offers possibilities that extend far beyond these initial explorations. MM platforms should enable real-time spatial manipulation of sound objects, facilitates multi-user collaborative composition, and supports dynamic acoustic environment modelling in ways previously unimaginable. These capabilities represent new dimensions of musical expression that composers are only beginning to explore.

### 4.4. The primacy of mixed reality over virtual environments

An important finding across all three groups was the consistent preference of Mixed Reality over purely virtual environments. This preference appears to stem from practitioners' deep understanding of the embodied nature of musical practice in composition, performance, or production. The skepticism about fully virtual performance environments, particularly from Classical Musicians and Composers, suggests that the physical connection to instruments and spaces remains fundamental to musical practice. This is in line with experimental results showing a consistent high perceived presence in Mixed Reality musical systems when remote participants are presented both as avatars or volumetric point clouds (e.g., Bruns et al. 2024; Schlagowski et al. 2023).

This suggests that the development of the MM should prioritize technologies that enhance rather than replace physical interaction with instruments and spaces.

### 4.5. Towards a Musical Metaverse compatible with musicians' needs

The features of the Musical Metaverse that we have identified as impacting dimensions such as Composition, Performance, and Education vary among the groups, with Composers and Producers focusing on space and technology (e.g. immersive studios, interfaces), while Classical Musicians focus on more practical aspects like monitoring and support from teachers.

All groups share an interest in learning simulators, but their infrastructure to support such simulators will differ: Producers might need high-end technology for digital twins, while composers should require other tools such as perceptual aids.

While Composers and Producers seem more willing to embrace technological advances, Classical Musicians may face challenges due to the need for teacher support. This reflects in different levels of external technological adoption and requirements. Performers' skepticism is present across groups, particularly among producers and composers. Overcoming this skepticism requires targeting specific stakeholders' concerns: for example, Classical Musicians may be more concerned about the emotional impact in live musical execution, while composers and producers may need to see the creative benefits in action.

Regarding Composers, success might look like an increase in the use of virtual interfaces for compositional experimentation. For Classical Musicians, overcoming skepticism might result in enhanced self-evaluation during virtual rehearsals. Producers, if successful, might create entire immersive production studios that fully integrate digital twins and malleable spaces for creative work.

### 4.5.1. Implications for design

These findings suggest several key considerations for the design of MM systems:

• Prioritize Mixed Reality approaches to maintain a better connection to physical instruments and spaces;

• Develop robust feedback systems that leverage the unique capabilities of virtual environments in terms of immersion and multisensory interactions;

• Focus initial development on educational applications;

• Design for space as a fundamental creative dimension;

• Ensure accessibility while maintaining high standards for audio quality and performance capability.

## 4.6. Limitations and future work

This study focused on three specific groups of music practitioners of two Italian music institutions. Future research should examine whether these findings generalize to other musical traditions and cultural contexts. Moreover, the studies have been conducted in an educational and academic setting that might have influenced the discussion. Therefore, further research should be conducted with musicians and stakeholders not involved in educational institutions.

Additionally, the use of the 'if by magic' framing device, while useful for encouraging creative thinking, may have led participants to envision capabilities beyond what is technically feasible in the near term. We deliberately chose not to provide participants with any experience of existing technologies and available applications to avoid influencing their responses. Future research should be devoted to generating more specific design guidelines by having users test and evaluate prototypes of MM instances that follow the results presented in this study.

## 5. Conclusion

The Musical Metaverse represents a potentially transformative technology for musical practice, but its development must be guided by careful consideration of the needs and concerns of different musical practitioners. This study suggests that while there is significant enthusiasm for certain applications, particularly in education and preparation, there are also important reservations about its use in performance contexts. Future development of the MM should focus on applications that enhance rather than replace traditional musical practices, with particular attention to maintaining the essential physical and embodied aspects of musical creativity.

Future research should examine how these findings translate into specific technical requirements and explore how the identified needs can be met within current technological constraints. Additionally, longitudinal studies of early MM implementations could provide valuable insights into how these technologies actually impact musical practice over time.

## Notes

1. First International Workshop on the Musical Metaverse, IEEE IS2 2024, Enrlangen, DE, 2024 (https://internetofsounds.net/1st-international-workshop-on-the-musical-metaverse/).
2. Sound of the Metaverse: Conference on Sound in Interactive Digital Environments, Chopin University of Music, Warsaw, PL, 2024 (https://soundsofthemetaverse.pl).
3. PatchWorld (https://patchxr.com).
4. TribeXR (https://www.tribexr.com).

## Figures and Captions

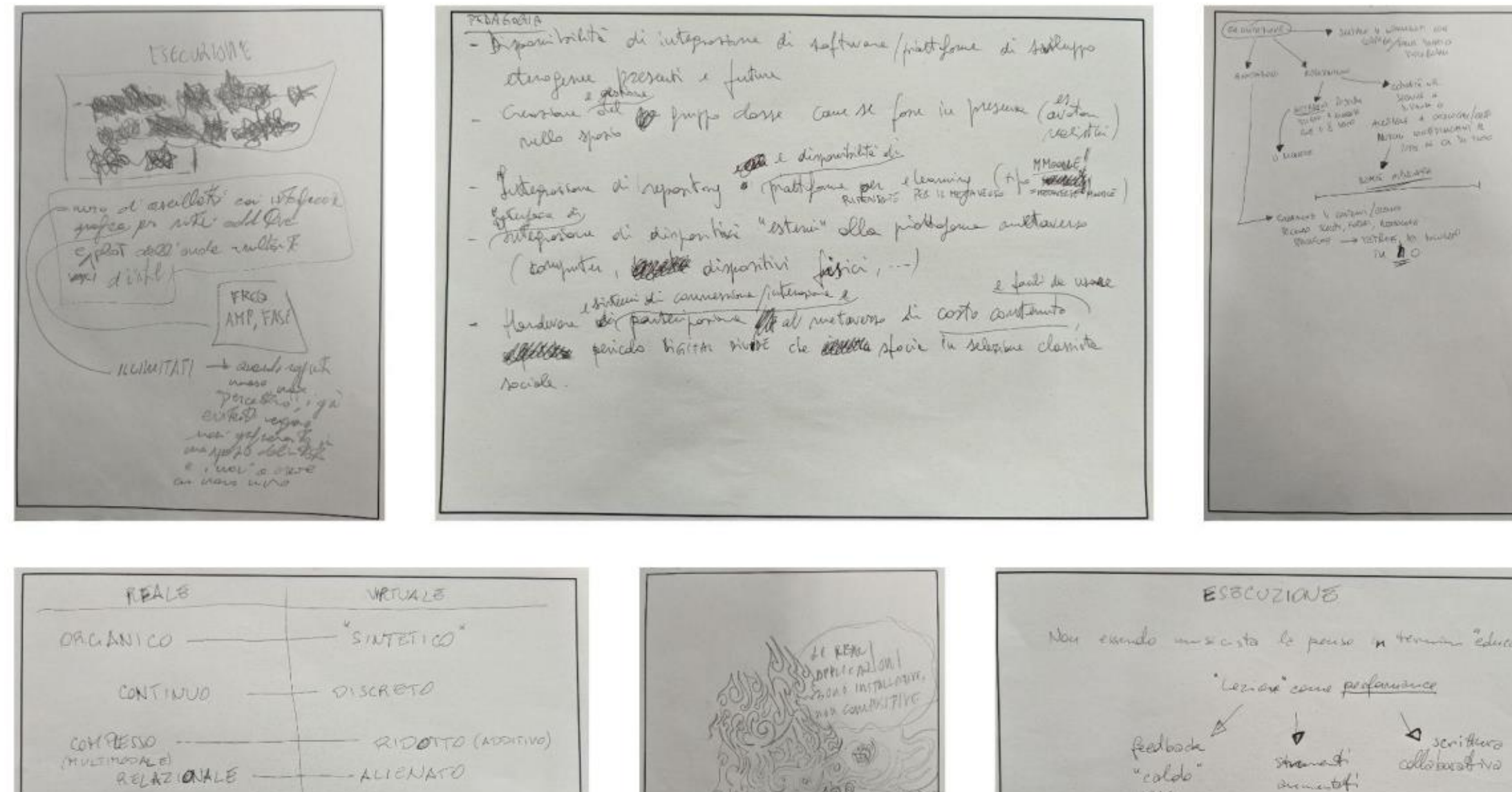

Figure 1 Caption: An example of the sketch sheets drawn by participants during the three workshops.

Figure 1 Alt Text: Collection of six frames arranged in two rows, showing the sketch sheets from workshop participants.

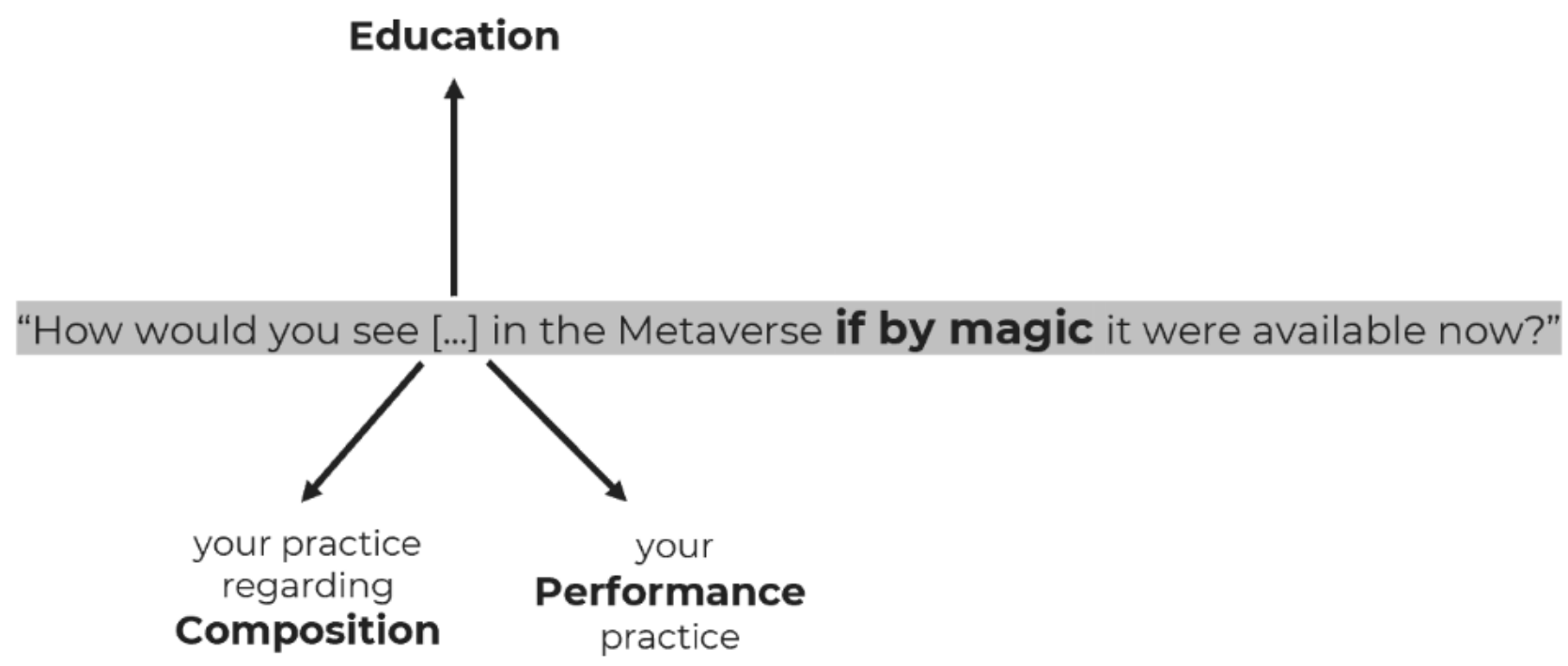

Figure 2 Caption: It shows the “If by magic” question used to encourage participants to envision their practice within the MM.

Figure 2 Alt Text: Diagram showing the workshop prompt question 'How would you see [...] in the Metaverse if by magic it were available now?' branching into three areas: Education (top), Composition (bottom left), and Performance practice (bottom right).

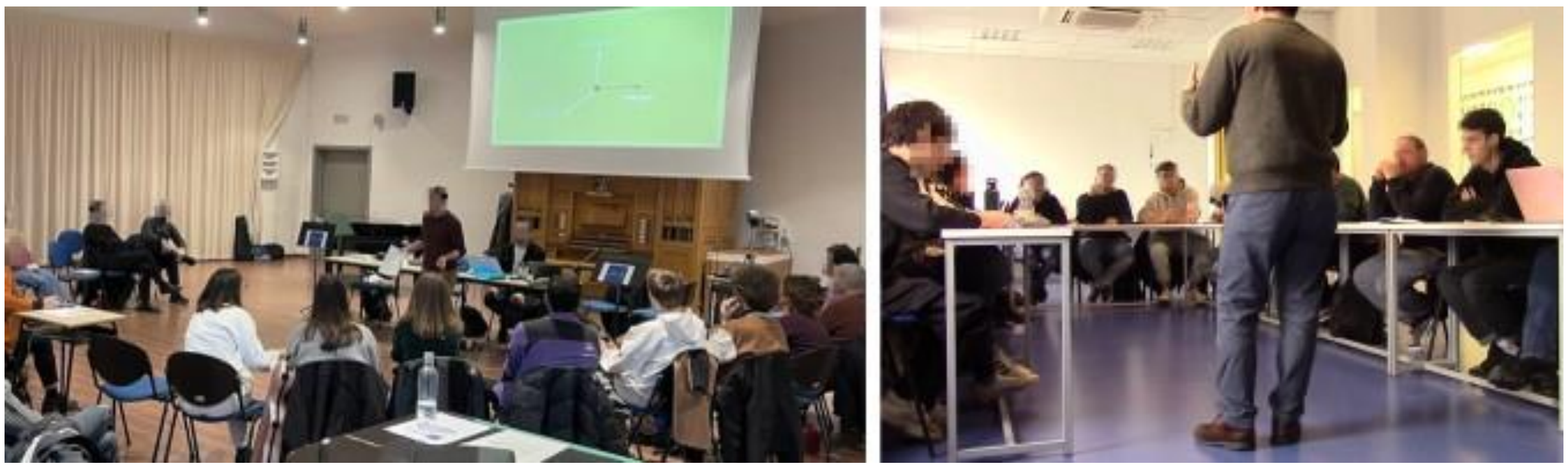

Figure 3 Caption: the facilitators explaining the basics of the MM. On the left, the workshop with Classical Musicians. On the right, the workshop with Producers and Sound Engineers.

Figure 3 Alt Text: Two photographs side by side showing workshop sessions. Left image shows a workshop with Classical Musicians, right image shows a workshop with Producers and Sound Engineers. Both show facilitators explaining Musical Metaverse concepts.

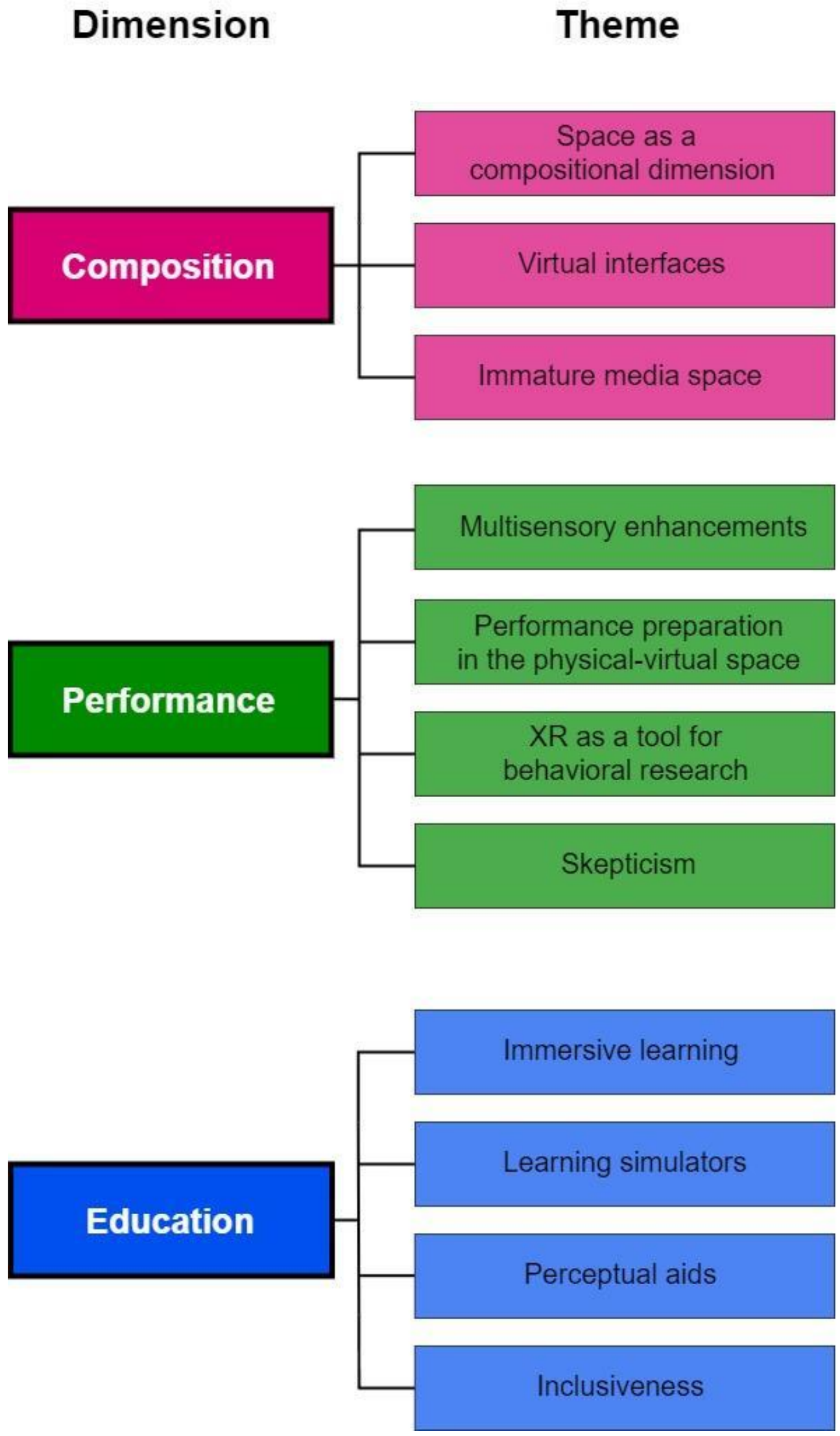

Figure 4 Caption: The themes for each dimension regarding the Electroacoustic Composers.

Figure 4 Alt Text: Hierarchical diagram showing themes for Electroacoustic Composers across three dimensions: Composition (pink), Performance (green), and Education (blue). Each dimension branches into multiple related themes.

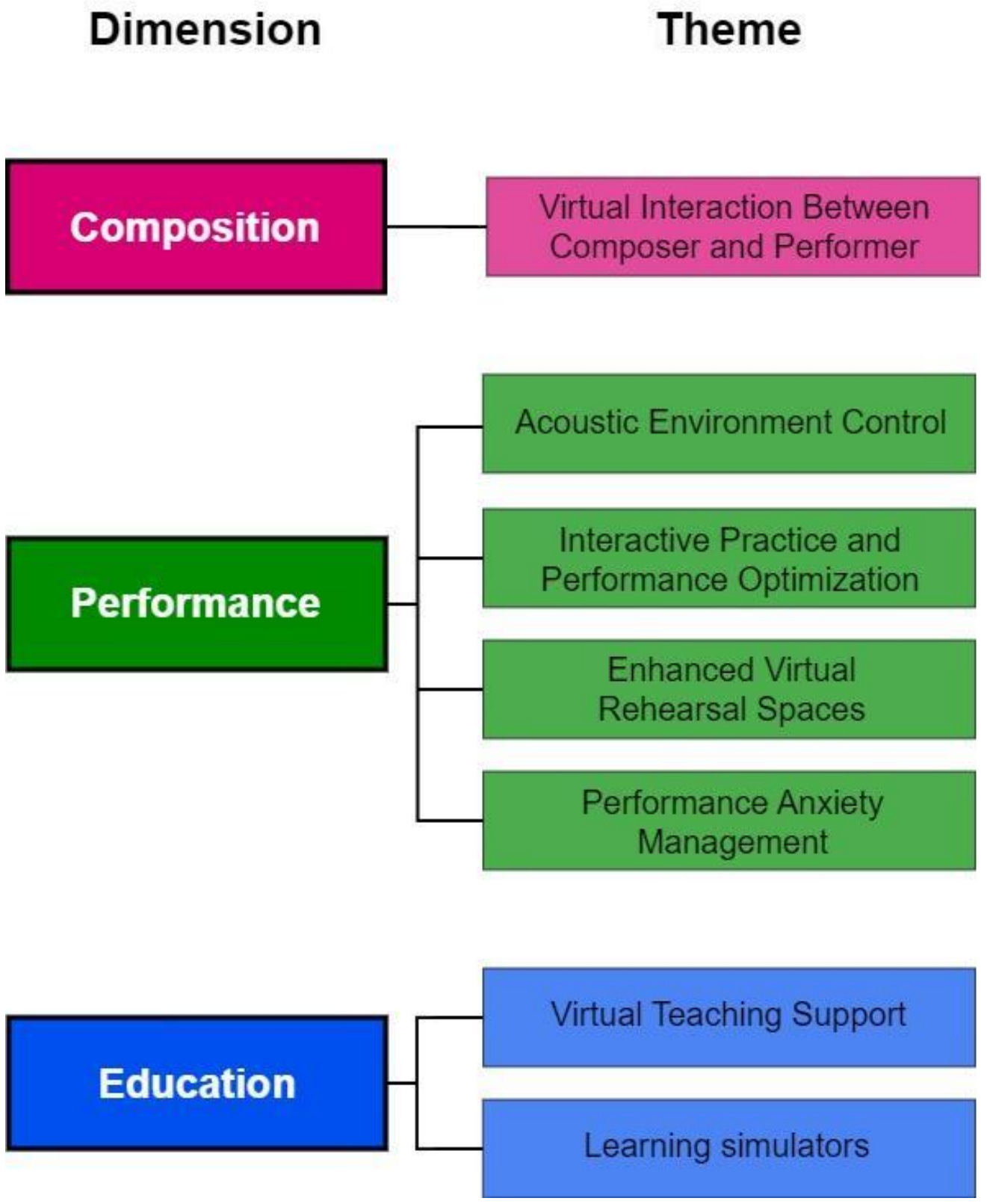

Figure 5 Caption: The themes for each dimension regarding the Classical Musicians.

Figure 5 Alt Text: Hierarchical diagram showing themes for Classical Musicians across three dimensions: Composition (single theme), Performance (multiple themes), and Education (two themes). Each section uses consistent color coding (pink, green, blue).

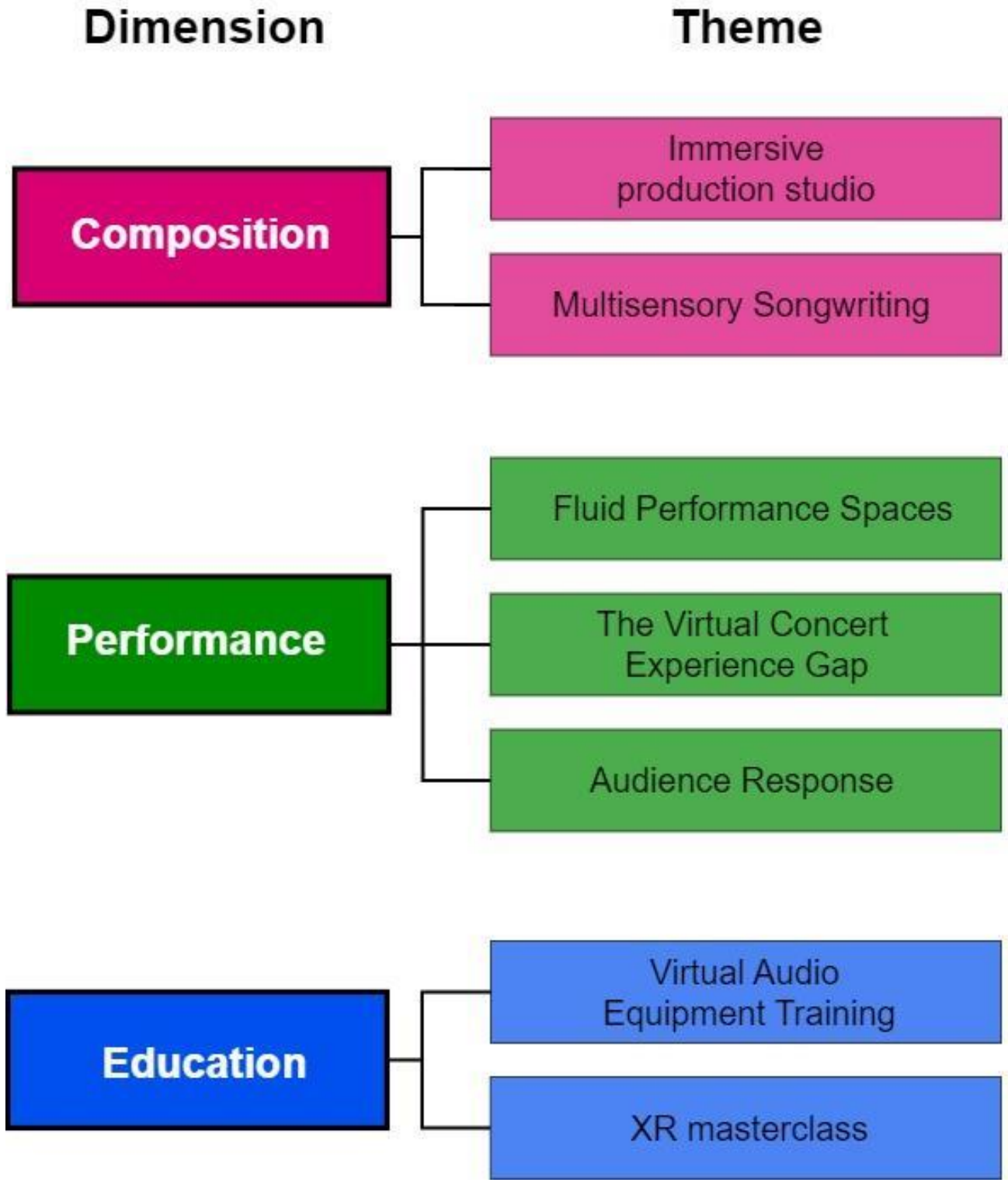

Figur 6 Caption: The themes for each dimension regarding Producers and Sound Engineers.

Figure 6 Alt Text: Hierarchical diagram showing themes for Producers and Sound Engineers across three dimensions: Composition (two themes), Performance (three themes), and Education (two themes). Uses consistent color coding (pink, green, blue).

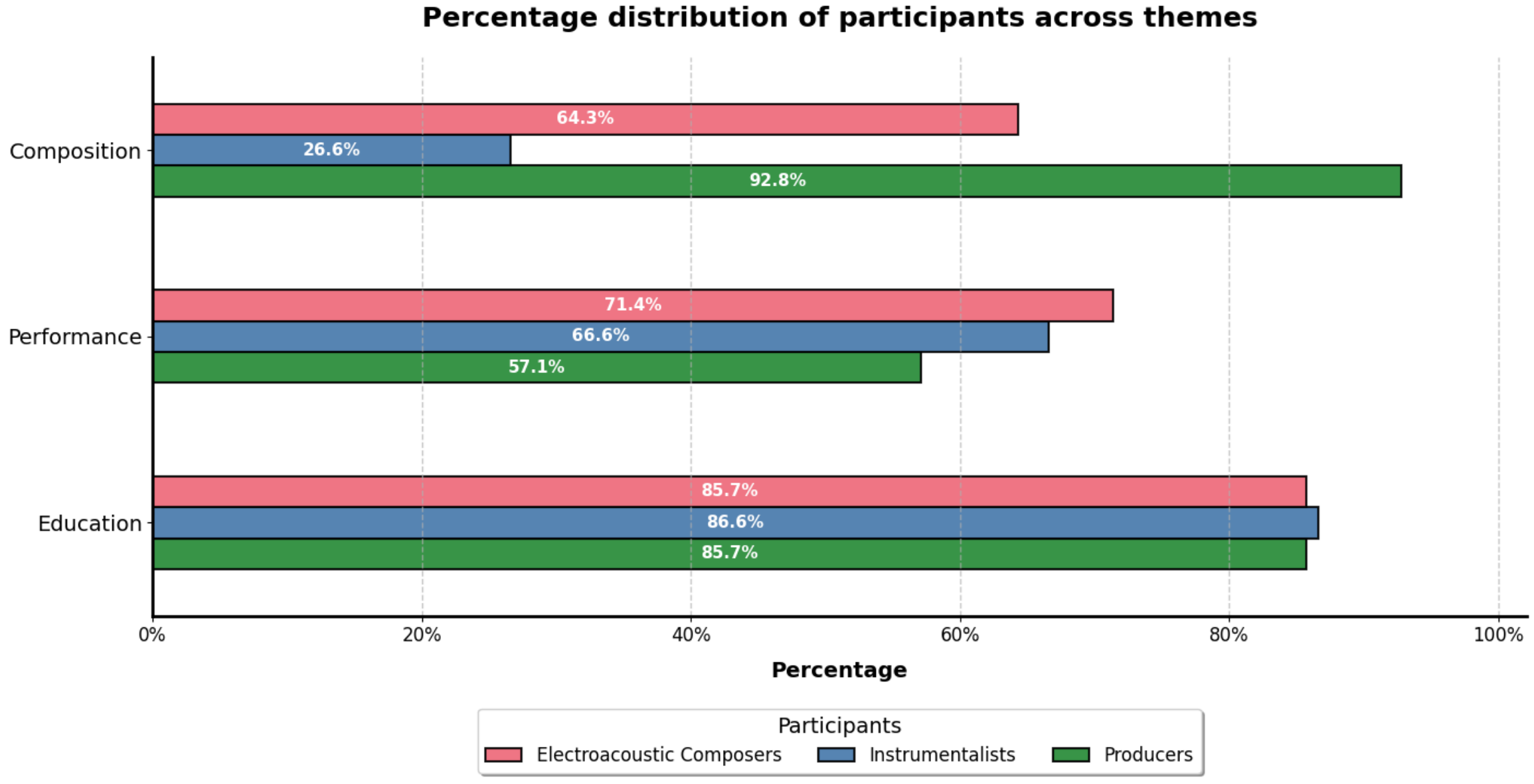

Figure 7 Caption: Percentage distribution of participants for each group across the three dimensions of the MM.

Figure 7 Alt Text: Bar chart comparing percentage distribution of participants across three dimensions (Composition, Performance, Education) for three groups: Electroacoustic Composers, Instrumentalists, and Producers. Shows higher engagement in Education across all groups.

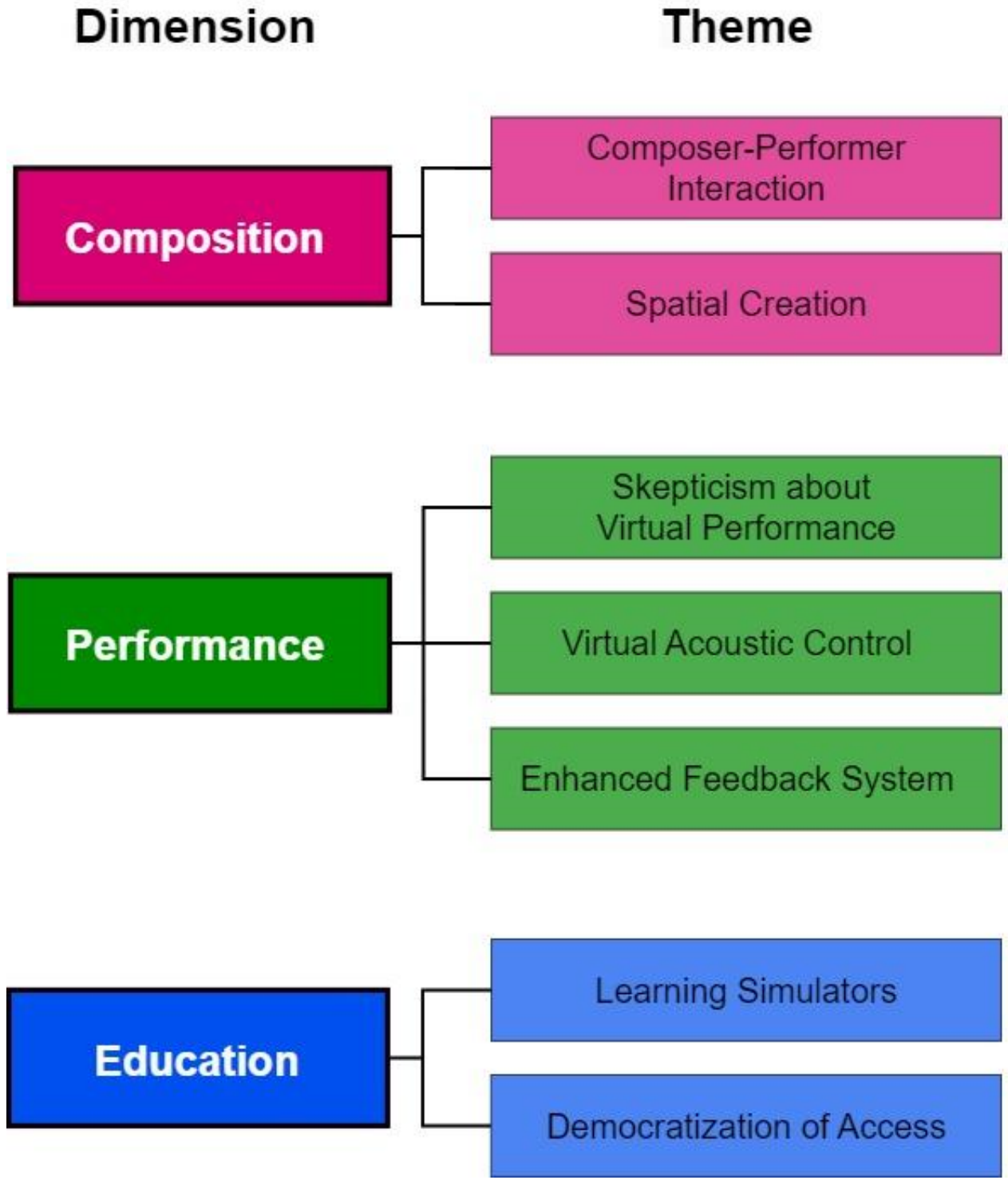

Figure 8 Caption: The themes for each dimension regarding the common theme across Electroacoustic composers, classical musicians, and producers and sound engineers.

Figure 8 Alt Text: Hierarchical diagram showing common themes across all three groups (Electroacoustic composers, classical musicians, and producers) organized by dimension: Composition (two themes), Performance (three themes), and Education (two themes). Uses consistent color coding.

**Acknowledgements**

Views and opinions expressed are however those of the author(s) only and do not necessarily reflect those of the European Union or the European Innovation Council. Neither the European Union nor the European Innovation Council can be held responsible for them.

**Disclosure statement**

No potential conflict of interest was reported by the author(s).

**Funding**

The authors received support from the MUR PNRR PRIN 2022 grant, prot. no. 2022CZWWKP, funded by Next Generation EU. Luca Turchet also acknowledges the support of the MUSMET project funded by the EIC Pathfinder Open scheme of the European Commission (grant agreement n. 101184379). Views and opinions expressed are however those of the author(s) only and do not necessarily reflect those of the European Union or the European Innovation Council. Neither the European Union nor the European Innovation Council can be held responsible for them.

**Notes on contributors**

**Alberto Boem** is currently a postdoctoral researcher at the Department of Information Engineering and Computer Science of the University of Trento, Italy. In 2010, he received his master's degree in 'Languages and Technologies of New Media' from the University of Udine, and in 2014, a Master of Arts in 'Interface Culture" from the University of Art and Design Linz. He received his Ph.D. in 'Human Informatics" (2019) from the University of Tsukuba. His research and artistic work focuses on interactive, embodied, and immersive systems. During the years, he explored areas such as deformable and shape-changing interfaces, multi-sensory interactions, and musical XR. His research activities have been presented in venues such as ACM DIS, IEEE VR, IS2, and NIME. His artistic work has been presented at several international venues such as Ars Electronica Festival, Sónar + D, STEIM, YCAM, MNAC, and the Guthman Musical Instrument Competition.

**Matteo Tomasetti** is a Professor of Electronic Music at the Conservatory of Music of Pescara, Italy. In 2024 he completed his PhD at the Department of Information Engineering and Computer Science of the University of Trento, Italy. He works with sound art, acousmatic and multimedia composition, interactive installation, live performance, and his research interests span spatial audio, extended reality, computer music composition and performance, and New Interfaces for Musical Expression (NIME). He is also a producer and live performer, working across various electronic music genres. Some of his works and performances have been presented at venues and festivals worldwide, including Ars Electronica, Venice Biennale, International Computer Music Conference, Gaudeamus Festival, New York City Electroacoustic

Music Festival, Videocittà Festival, and Live Cinema Festival. His research has been published in scientific journals such as IEEE, Springer, and Taylor & Francis, and presented at various conferences, such as NIME, the International Conference on Immersive and 3D Audio, and the IEEE International Symposium on the Internet of Sounds.

**Alessio Gabriele** is a composer, sound and multimedia artist, violinist, and computer scientist, he graduated from the Conservatory of L'Aquila in Violin (2001) and Electronic Music (2013) and holds bachelor's (2002) and master's degrees (2009) in Computer Science from the University of L'Aquila. He also pursued advanced studies in Composition and Music Performance. Appointed a full-time Professor in Multimedia in 2020, he previously taught at conservatories in L'Aquila, Salerno, Cuneo, and Bari and now holds a tenure-track position at the Conservatory 'A. Casella' in L'Aquila. His work explores the interplay between music production, embodied cognition, multimedia, and ICT, integrating artistic expression with technology. As a composer, he creates acousmatic, instrumental, and mixed music, develops augmented instruments, and contributes to interactive operas, sound art installations, and adaptive music systems. His works have been featured in prestigious events such as Accademia Nazionale di Santa Cecilia, NYCEMF, and Accademia Filarmonica Romana. Collaborating with CRM Centro Ricerche Musicali and other artists, he has realized permanent sound art installations and multimedia projects. His research includes publications with MIT Press and presentations at international conferences. As a violinist, he has performed with renowned orchestras and chamber ensembles in Italy and abroad since 1998.

**Agostino di Scipio** is a composer, sound artist, and scholar. Graduated in Composition and Electronic Music from the Conservatory of L'Aquila, appointed PhD at the University Paris VIII with a practice-led research on the notion of 'liveness' in highly mediatized performance. His creative work focuses on original and peculiar techniques in sound generation and transmission, often exploring 'man-machine-environment' networks of sonic interactions. He is currently a professor of Electroacoustic Composition at the Conservatory of L'Aquila. He was artist-in-residence at world-wide institutions such as DAAD Künstlerprogramm (Berlin, 2004–2005), the Warszaw Biennale (2019) and many others. His output as a composer and sound artist was the focus of various publications, including a special issue of Contemporary Music Review and a Computer Music Journal interview. Author of several articles and essays appeared in international peer-reviewed publications, he published textbooks and monographs such Circuiti del tempo. Un percorso storico-critico sulla creatività musicale elettroacustica e informatica (LIM 2020). He serves as a guest editor for the Journal of New Music Research and Music/Technology.

**Alessandra Micalizzi** (PhD in Communication and New Technologies) is an Associate Professor at Pegaso University and a Senior Lecturer at SAE Institute in Milan. She coordinates the PhD programme in Digital Humanities and leads the Center for Research in Digital Humanities at Pegaso University. Following her post-doctoral fellowship, she collaborated with several academic institutions, including IULM University, IUSTo, IUSVe, IED, and LIMEC. Her research interests focus on the impact of new technologies on everyday practices. She has explored emotions and the role of digital spaces in fostering new forms of emotional sharing. Recently, her work has concentrated on gender studies within cultural industries and the influence of Artificial Intelligence on creative production. Among her recent works, she co-authored 'Video Game Therapy' (UTET Università); 'Artificial Creativity. Looking at the future of Digital Culture' (Springer Nature).

**Luca Turchet** is an Associate Professor at the Department of Information Engineering and Computer Science, University of Trento, Italy. He holds a master's degree in Computer Science (2006) from the University of Verona, degrees in classical guitar (2007) and composition (2009) from the Music Conservatory of Verona, as well as in electronic music (2015) from the Royal College of Music, Stockholm. He earned his Ph.D. in Media Technology (2013) from Aalborg University Copenhagen. His research was supported by international funding agencies such as the European Commission and the European Space Agency. He co-founded the company Elk. He is the Chair of the IEEE Emerging Technology Initiative on the Internet of Sounds and the founding president of the Internet of Sounds Research Network. He serves as an Associate Editor for the Journal of the Audio Engineering Society and IEEE Transactions on Human-Machine Systems.

## References

Andersen, K., and R. Wakkary. 2019. "The Magic Machine Workshops: Making Personal Design Knowledge." In Proceedings of the 2019 CHI Conference on Human Factors in Computing Systems, 1–13. New York, NY: ACM.

Arslan, C., F. Berthaut, A. Beuchey, P. Cambourian, and A. Paté. 2022, June. "Vibrating Shapes: Design and Evolution of a Spatial Augmented Reality Interface for Actuated Instruments." In NIME 2022. PubPub.

Azubuike, O. B., O. Adegboye, and H. Quadri. 2021. "Who Gets to Learn in a Pandemic? Exploring the Digital Divide in Remote Learning during the COVID-19 Pandemic in Nigeria." International Journal of Educational Research Open 2:100022.

Ball, M. 2024. The Metaverse: Building the Spatial Internet. New York, NY: Liveright Pub Corp.

Basanta, A. 2015. "Extending Musical Form Outwards in Space and Time: Compositional Strategies in Sound Art and Audiovisual Installations." Organised Sound 20 (2): 171–181. https://doi.org/10.1017/S1355771815000059.

Bell, J. 2023. "Networked Music Performance in PatchXR and FluCoMa." International computer music conference (ICMC) 2023.

Benjamins, R., Y. R. Viñuela, and C. Alonso. 2023. "Social and Ethical Challenges of the Metaverse: Opening the Debate." AI and Ethics 3 (3): 689–697. https://doi.org/10.1007/s43681-023-00278-5.

Boem, A., M. Tomasetti, A. Gabriele, A. Di Scipio, and L. Turchet. 2024. "User Needs in the Musical Metaverse: A Case Study with Electroacoustic Musicians." Proceedings of the International Conference on New Interfaces for Musical Expression.

Boem, A., M. Tomasetti, and L. Turchet. 2024. "Harmonizing the Musical Metaverse: Unveiling Needs, Tools, and Challenges from Experts' Point of View." Proceedings of the International Conference on New Interfaces for Musical Expression.

Boem, A., M. Tomasetti, and L. Turchet. 2025. "Issues and Challenges of Audio Technologies for the Musical Metaverse." Journal of the Audio Engineering Society 73 (3): 94–114. https://doi.org/10.17743/jaes.2022.0193.

Braun, V., and V. Clarke. 2019. "Reflecting on Reflexive Thematic Analysis." Qualitative Research in Sport, Exercise and Health 11 (4): 589–597. https://doi.org/10.1080/2159676X.2019.1628806.

Bruns, L., B. Saurbier, T. M. Voong, and M. Oehler. 2024. "Presence and Flow in Virtual and Mixed Realities for Music-related Educational Settings." In 2024 IEEE 5th International Symposium on the Internet of Sounds (IS2), 1–7. IEEE.

Cheng, R., N. Wu, S. Chen, and B. Han. 2022. "Will Metaverse Be Nextg Internet? Vision, Hype, and Reality." IEEE Network 36 (5): 197–204. https://doi.org/10.1109/MNET.117.2200055.

Cheng, R., N. Wu, M. Varvello, S. Chen, and B. Han. 2022. "Are We Ready for Metaverse? A Measurement Study of Social Virtual Reality Platforms." In Proceedings of the 22nd ACM Internet Measurement Conference, 504–518. New York, NY: ACM.

Chengoden, R., N. Victor, T. Huynh-The, G. Yenduri, R. H. Jhaveri, M. Alazab, and T. R. Gadekallu. 2023. "Metaverse for Healthcare: A Survey on Potential Applications, Challenges and Future Directions." IEEE Access 11:12765–12795. https://doi.org/10.1109/ACCESS.2023.3241628.

Churchill, E. F., and D. Snowdon. 1998. "Collaborative Virtual Environments: An Introductory Review of Issues and Systems." Virtual Reality 3 (1): 3–15. https://doi.org/10.1007/BF01409793.

Ciciliani, M. 2020. "Virtual 3D Environments as Composition and Performance Spaces." Journal of New Music Research 49 (1): 104–113. https://doi.org/10.1080/09298215.2019.1703013.

Deines, E., F. Michel, M. Bertram, J. Mohring, and H. Hagen. 2007. "Simulation, Visualization, and Virtual Reality Based Modeling of Room Acoustics".

Di Scipio, A. 2000. "The Technology of Musical Experience in the 20th Century." Rivista Italiana di Musicologia 35 (1/2): 247–275.

Di Scipio, A. 2011. "Listening to Yourself through the Otherself: On Background Noise Study and Other Works." Organised Sound 16 (2): 97–108. https://doi.org/10.1017/S1355771811000033.

Dwivedi, Y. K., L. Hughes, A. M. Baabdullah, S. Ribeiro-Navarrete, M. Giannakis, M. M. Al-Debei, Denis Dennehy, et al. 2022. "Metaverse beyond the Hype: Multidisciplinary Perspectives on Emerging Challenges, Opportunities, and Agenda for Research, Practice and Policy." International Journal of Information Management 66:102542. https://doi.org/10.1016/j.ijinfomgt.2022.102542.

Dziwis, D., and H. von Coler. 2023. "The Entanglement: Volumetric Music Performances in a Virtual Metaverse Environment." Journal of Network Music and Arts 5 (1): 3.

El Saddik, A., F. Lamberti, S. Mann, F. G. Pratticò, R. Thawonmas, and Y. Yuan. 2024. "Metaverse and eXtended uniVerse (XV): Opportunities and Challenges for Consumer Technologies." IEEE Consumer Electronics Magazine.

Gerry, L., S. Dahl, and S. Serafin. 2019. "ADEPT: Exploring the Design, Pedagogy, and Analysis of a Mixed Reality Application for Piano Training." 16th Sound and Music Computing Conference. Sound and Music Computing Network.

Glaser, B., and A. Strauss. 2017. Discovery of Grounded Theory: Strategies for Qualitative Research. New York, NY: Routledge.

Gómez-Sirvent, J. L., F. López de la Rosa, R. Sánchez-Reolid, R. Morales Herrera, and A. Fernández-Caballero. 2024. "Musical Instruments in Extended Reality: A Systematic Review." International Journal of Human – Computer Interaction 1–20. https://doi.org/10.1080/10447318.2024.2431352.

Gursoy, D., S. Malodia, and A. Dhir. 2022. "The Metaverse in the Hospitality and Tourism Industry: An Overview of Current Trends and Future Research Directions." Journal of Hospitality Marketing & Management 31 (5): 527–534. https://doi.org/10.1080/19368623.2022.2072504.

Hamilton, R. 2019. "Collaborative and Competitive Futures for Virtual Reality Music and Sound." In 2019 IEEE Conference on Virtual Reality and 3D User Interfaces (VR), 1510–1512. Osaka, Japan: IEEE.

Hussein, I., M. Mahmud, and A. O. M. Tap. 2014. "A Survey of User Experience Practice: A Point of Meet between Academic and Industry." 2014 3rd International Conference on User Science and Engineering (i-USEr), IEEE.

Inoue, A., Y. Ikeda, K. Yatabe, and Y. Oikawa. 2019. "Visualization System for Sound Field Using see-Through Head-Mounted Display." Acoustical Science and Technology 40 (1): 1–11. https://doi.org/10.1250/ast.40.1.

Johnson, D., D. Damian, and G. Tzanetakis. 2020. "Evaluating the Effectiveness of Mixed Reality Music Instrument Learning with the Theremin." Virtual Reality 24 (2): 303–317. https://doi.org/10.1007/s10055-019-00388-8.

Lepri, G., and A. McPherson. 2019. "Making up Instruments: Design Fiction for Value Discovery in Communities of Musical Practice." In Proceedings of the 2019 on Designing Interactive Systems Conference, 1–13. New York, NT: ACM.

Lin, H., S. Wan, W. Gan, J. Chen, and H. C. Chao. 2022. "Metaverse in Education: Vision, Opportunities, and Challenges." In 2022 IEEE International Conference on Big Data (Big Data), 2857–2866. Osaka, Japan: IEEE.

Lindborg, P., S. S. Chopra, and K. Groß-Vogt. 2023. "Editorial: Data Perceptualization for Climate Science Communication." Frontiers in Psychology 14:1263971. https://doi.org/10.3389/fpsyg.2023.1263971.

Lombardo, V., A. Valle, J. Fitch, K. Tazelaar, S. Weinzierl, and W. Borczyk. 2009. "A Virtual-Reality Reconstruction of Poeme Electronique Based on Philological Research." Computer Music Journal 33 (2): 24–47. https://doi.org/10.1162/comj.2009.33.2.24.

Lopes, F., and C. Guedes. 2020. "Composing Music with a Space." Perspectives of New Music 58 (1): 5–22.

Loveridge, B. 2024. "Key Considerations for Duo Singing in Virtual Reality and Videoconferencing: An Exploratory Study with Bigscreen and Zoom." 2024 IEEE 5th International Symposium on the Internet of Sounds (IS2), IEEE.

Macedo, F. 2015. "Investigating Sound in Space: Five Meanings of Space in Music and Sound Art." Organised Sound 20 (2): 241–248. https://doi.org/10.1017/S1355771815000126.

Makransky, G., and G. B. Petersen. 2021. "The Cognitive Affective Model of Immersive Learning (CAMIL): A Theoretical Research-based Model of Learning in Immersive Virtual Reality." Educational Psychology Review 33 (3): 937–958. https://doi.org/10.1007/s10648-020-09586-2.

Martín, G. F. 2018. "Social and Psychological Impact of Musical Collective Creative Processes in Virtual Environments; the Avatar Orchestra Metaverse in Second Life." Music Technology 75:75–87.

Milgram, P., and F. Kishino. 1994. "A Taxonomy of Mixed Reality Visual Displays." IEICE Transactions on Information and Systems 77 (12): 1321–1329.

Morreale, F., S. A. Bin, A. P. McPherson, P. Stapleton, and M. Wanderley. 2020. "A NIME of the Times: Developing an Outward-looking Political Agenda for This Community." In International Conference on New Interfaces for Musical Expression 2020, 191–197. Birmingham: NIME.

Morreale, Fabio, Nicolas Gold, Cécile Chevalier, and Raul Masu. 2023. NIME Principles & Code of Practice on Ethical Research (1.1).

Munn, N., and D. Weijers. 2023. "The Real Ethical Problem with Metaverses." Frontiers in Human Dynamics 5:1226848. https://doi.org/10.3389/fhumd.2023.1226848.

Nijs, L., and B. Behzadaval. 2024. "Laying the Foundation for Augmented Reality in Music Education." IEEE Access 12: 100628–100645.

Oh, C. S., J. N. Bailenson, and G. F. Welch. 2018. "A Systematic Review of Social Presence: Definition, Antecedents, and Implications." Frontiers in Robotics and AI 5:409295.

Onderdijk, K. E., L. Bouckaert, E. Van Dyck, and P. J. Maes. 2023. "Concert Experiences in Virtual Reality Environments." Virtual Reality 27 (3): 2383–2396. https://doi.org/10.1007/s10055-023-00814-y.

Orman, E. K., H. E. Price, and C. R. Russell. 2017. "Feasibility of Using an Augmented Immersive Virtual Reality Learning Environment to Enhance Music Conducting Skills." Journal of Music Teacher Education 27 (1): 24–35. https://doi.org/10.1177/1057083717697962.

Ørngreen, R., and K. T. Levinsen. 2017. "Workshops as a Research Methodology." Electronic Journal of E-Learning 15 (1): 70–81.

Park, J., Y. Choi, and K. M. Lee. 2024. "Research Trends in Virtual Reality Music Concert Technology: A Systematic Literature Review." IEEE Transactions on Visualization and Computer Graphics 30 (5): 2195–2205.

Park, S. M., and Y. G. Kim. 2022. "A Metaverse: Taxonomy, Components, Applications, and Open Challenges." IEEE Access 10:4209–4251. https://doi.org/10.1109/ACCESS.2021.3140175.

Paterson, J., and H. Lee, eds. 2021. 3D Audio. London: Routledge.

Petersen, G. B., G. Petkakis, and G. Makransky. 2022. "A Study of How Immersion and Interactivity Drive VR Learning." Computers & Education 179:104429. https://doi.org/10.1016/j.compedu.2021.104429.

Pirchner, A. 2020. "Ergodic and Emergent Qualities of Realtime Scores. Anna and Marie and Gamified Audiovisual Compositions." In Proceedings of the International Conference on Technologies for Music Notation and Representation–TENOR, Hamburg, Germany, Vol. 20, 189–197.

Ppali, S., M. Scorer, E. Ppali, B. Branch, and A. Covaci. 2024. "Remote Rhythms: Audience-Informed Insights for Designing Remote Music Performances." In Proceedings of the 2024 ACM Designing Interactive Systems Conference, 2675–2690. New York, NY: ACM.

Renaud, A., A. Carôt, and P. Rebelo. 2007. "Networked Music Performance: State of the art." Proceedings of the AES 30th International Conference, Saariselkä: Finland.

Ritterbusch, G. D., and M. R. Teichmann. 2023. "Defining the Metaverse: A Systematic Literature Review." IEEE Access 11:12368–12377. https://doi.org/10.1109/ACCESS.2023.3241809.

Sai, S., A. Garg, and V. Chamola. 2024. "Navigating the Metaverse: A Comprehensive Analysis of Consumer Electronics Prospects and Challenges." ACM Transactions on Internet Technology. https://doi.org/10.1145/3680545.

Schlagowski, R., D. Nazarenko, Y. Can, K. Gupta, S. Mertes, M. Billinghurst, and E. André. 2023. "Wish You Were Here: Mental and Physiological Effects of Remote Music Collaboration in Mixed Reality." In Proceedings of the 2023 CHI Conference on Human Factors in Computing Systems, 1–16. New York, NY: ACM.

Schroeder, R., A. Steed, A. S. Axelsson, I. Heldal, Å Abelin, J. Wideström, A. Nilsson, and M. Slater. 2001. "Collaborating in Networked Immersive Spaces: As Good as Being There Together?" Computers & Graphics 25 (5): 781–788. https://doi.org/10.1016/S0097-8493(01)00120-0.

Serafin, S., A. Adjorlu, N. Nilsson, L. Thomsen, and R. Nordahl. 2017. "Considerations on the use of Virtual and Augmented Reality Technologies in Music Education." In 2017 IEEE Virtual Reality Workshop on K-12 Embodied Learning through Virtual & Augmented Reality (KELVAR), 1–4. Los Angeles, CA: IEEE.

Serafin, S., C. Erkut, J. Kojs, N. C. Nilsson, and R. Nordahl. 2016. "Virtual Reality Musical Instruments: State of the art, Design Principles, and Future Directions." Computer Music Journal 40 (3): 22–40. https://doi.org/10.1162/COMJ_a_00372.

Shekhar, S., S. K. Feiner, and W. G. Aref. 2015. "Spatial Computing." Communications of the ACM 59 (1): 72–81. https://doi.org/10.1145/2756547.

Spence, E. H. 2008. "Meta Ethics for the Metaverse: The Ethics of Virtual Worlds." Current Issues in Computing and Philosophy 175 (3): 3–12.

Stephenson, N. 1994. Snow Crash. London: Penguin.

Strömberg, H., I. Pettersson, J. Andersson, A. Rydström, D. Dey, M. Klingegård, and J. Forlizzi. 2018. "Designing for Social Experiences with and within Autonomous Vehicles – Exploring Methodological Directions." Design Science 4:e13. https://doi.org/10.1017/dsj.2018.9.

Tang, F., X. Chen, M. Zhao, and N. Kato. 2023. "The Roadmap of Communication and Networking in 6G for the Metaverse." IEEE Wireless Communications 30 (4): 72–81. https://doi.org/10.1109/MWC.019.2100721.

Turchet, L. 2023. "Musical Metaverse: Vision, Opportunities, and Challenges." Personal and Ubiquitous Computing 27 (5): 1811–1827. https://doi.org/10.1007/s00779-023-01708-1.

Turchet, L., C. Fischione, G. Essl, D. Keller, and M. Barthet. 2018. "Internet of Musical Things: Vision and Challenges." IEEE Access 6:61994–62017. https://doi.org/10.1109/ACCESS.2018.2872625.

Turchet, L., R. Hamilton, and A. Çamci. 2021. "Music in Extended Realities." IEEE Access 9:15810–15832. https://doi.org/10.1109/ACCESS.2021.3052931.

Vigevano, L., and P. Mattei. 2025. "The Challenges of Distance Learning in Italy: New Inequalities and Implications for Inclusive Education." International Journal of Inclusive Education 29 (8): 1308–1322.

Wang, H., H. Ning, Y. Lin, W. Wang, S. Dhelim, F. Farha, and M. Daneshmand. 2023. "A Survey on the Metaverse: The State-of-the-art, Technologies, Applications, and Challenges." IEEE Internet of Things Journal 10 (16): 14671–14688. https://doi.org/10.1109/JIOT.2023.3278329.

Wang, Y., Z. Su, N. Zhang, R. Xing, D. Liu, T. H. Luan, and X. Shen. 2023. "A Survey on Metaverse: Fundamentals, Security, and Privacy." IEEE Communications Surveys & Tutorials 25 (1): 319–352. https://doi.org/10.1109/COMST.2022.3202047.

Waters, R. C., and J. W. Barrus. 1997. "The Rise of Shared Virtual Environments." IEEE Spectrum 34 (3): 20–25. https://doi.org/10.1109/6.576004.

Young, G. W. 2024. "Ethical Considerations in the Production and Consumption of Music in the Metaverse." In 2024 IEEE 5th International Symposium on the Internet of Sounds (IS2), 1–9. Erlangen, DE: IEEE.

Zallio, M., and P. J. Clarkson. 2022. "Designing the Metaverse: A Study on Inclusion, Diversity, Equity, Accessibility and Safety for Digital Immersive Environments." Telematics and Informatics 75:101909. https://doi.org/10.1016/j.tele.2022.101909.

Zhang, H. 2024. "Metaverse VR Technologies in Contemporary Chinese Music Education." Interactive Learning Environments 33 (1): 821–836.